\documentclass[a4paper,12pt]{article}
\usepackage{jheppub,booktabs}
\usepackage{amsmath,amssymb}
\usepackage{xcolor,colortbl}
\usepackage{graphicx}
\usepackage{hepunits}
\usepackage{enumerate}
\usepackage[utf8x]{inputenc}
\usepackage{tikz-feynman}
\usepackage{comment}
\usepackage{bbm}
\usepackage{bbold}
\usepackage{xcolor}

\preprint{}
\author{Simone Blasi and}
\author{Florian Goertz}
\affiliation{Max-Planck-Institut f{\"u}r Kernphysik\\ Saupfercheckweg 1, 69117 Heidelberg, Germany}
\emailAdd{simone.blasi@mpi-hd.mpg.de}
\emailAdd{fgoertz@mpi-hd.mpg.de}

\title{Softened Goldstone-Symmetry Breaking}

\abstract{We propose a new way of breaking the Goldstone symmetry in composite Higgs models,
restoring the global symmetry in the mixings between the elementary and composite fermions
by completing the former to full representations of this symmetry.
The Goldstone symmetry is in turn broken softly by vector-like mass terms in the {\it elementary} sector only.
The resulting softened explicit breaking allows for a light Higgs boson, as found at the LHC, and a heavy top
quark, without the need of light top partners around the Goldstone scale $f \sim {\rm TeV} \ll m_{\rm comp.}$,
which remain elusive at the LHC,
while we recover the standard scenario in the limit of infinite vector-like masses.}
\date{\today}

\begin{document}
\maketitle

\section{Introduction}\label{sec:int}
Composite Higgs (CH) models offer a promising means to solve the hierarchy problem since the
Higgs boson is no longer a fundamental scalar but rather a bound state of a new strong interaction, that can be resolved above the TeV scale, and its mass is thus saturated in the IR \cite{Kaplan:1983fs,Kaplan:1983sm,Dugan:1984hq}. 
Moreover, the conventional assumption of the Higgs being a (pseudo-)Goldstone boson of a spontaneously broken global symmetry ($SO(5)\!\to \!SO(4)$ in minimal models) provides a reasoning for its lightness compared to other new states. In the same framework, partially composite fermions (elementary fields mixing linearly with the composite sector of bound states) can also explain the hierarchical structure of fermion masses and mixings \cite{Kaplan:1991dc,Agashe:2004rs,Contino:2003ve,Contino:2006qr} and provide a dynamical origin for EWSB, mostly triggered by the large top-quark compositeness. The latter explicitly breaks the global symmetry, since the Standard Model (SM) fermions do not fill complete representations of the global symmetry that could couple to the composite sector in an invariant way, and thereby induces a potential for the Goldstone Higgs, intertwining flavor and EWSB.

Minimal models are however already in tension 
with the absence of {\it light top partners} at the LHC, which are required to keep the Higgs light by reducing the Goldstone-symmetry breaking \cite{Contino:2006qr,Matsedonskyi:2012ym,Pomarol:2012qf,Csaki:2008zd,DeCurtis:2011yx}, threatening the viability of explicit CH incarnations (see, e.g., \cite{Goertz:2018dyw}).
While one possibility to avoid this is the inclusion of a realistic lepton sector generating small neutrino masses, which can have a larger impact on the Higgs potential than naively expected \cite{Carmona:2014iwa} with interesting consequences for flavor physics \cite{Carmona:2015ena,Carmona:2016mjr,Carmona:2017fsn,Carmona:2014iwa,Goertz:2018dyw} \footnote{See also \cite{Panico:2012uw} for a solution via an enlarged quark sector.}, here we want to explore 
an orthogonal solution, changing in fact the nature of the explicit Goldstone symmetry breaking.

Indeed, the latter could be significantly reduced if the SM fermions would be uplifted to complete representations of the global symmetry. In that case, their mixing with the composite sector, determining their degree of 'compositeness', would no longer violate the global symmetry. Since we did not observe additional light fermions so far, the symmetry still needs to be broken, which can however now be done by introducing vector-like mass terms for the new {\it elementary} fermions. This will shift the source of explicit breaking to a different sector and thereby corresponds to a fundamentally different approach of breaking the Goldstone symmetry. The breaking is now 'soft', since induced by mass terms in the elementary sector. Contrary to the conventional case, the underlying interactions between the SM fermions and the constituents of the composite states are now symmetry preserving.
In particular, the setup will lead to a different parameteric structure of the mass of the composite Higgs, with the potential to lift the light top partners. The purpose of this article is to work out the phenomenological and conceptual consequences of this approach.
Here, we focus on the minimal $SO(5)/SO(4)$ composite Higgs scenario \cite{Agashe:2004rs}, but the considerations
can easily be extended to different cosets.

This article is organized as follows.
In Sec.~\ref{sec:setup}, we will introduce the setup, including the field 
content and corresponding $SO(5)$ representations, 
and will discuss the nature of breaking of the Goldstone symmetry. In Sec.~\ref{sec:2site} we will confront 
our general discussion with explicit results obtained for
a two-site incarnation of the CH framework.
After that, in Sec.~\ref{sec:BM}, we will present 
benchmark spectra for our scenario of
soft (vector-like) Goldstone breaking (sGB), while in Sec.~\ref{sec:Gen} we demonstrate that raising the top partners is really related to a symmetry and not accidental.
Finally, we conclude in Sec.~\ref{sec:conc}.

\section{General Setup}

\label{sec:setup}
We consider just the third generation of quarks and lift the elementary fields to complete
$SO(5)$ multiplets, mixing linearly with the composite resonances 
$\tilde \Psi^T = U (Q,\,\tilde T)^T\,,$ which have been decomposed
into fourplets and singlets under the unbroken $SO(4)$ subgroup via the
CCWZ prescription, with $U$ the Goldstone matrix. The mixings with the resonances
now respect $SO(5)$, and its explicit breaking is sequestered
to the elementary sector.\footnote{
Note that the impact of the multiplets corresponding to the lighter generations
on the Higgs potential remains negliglible due to their small degree of compositeness.
Moreover, the corresponding elementary vector-like partners can lie much above the compositeness scale, effectively recovering the conventional
case.}

We recall that in the MCHM$_5$ \cite{Contino:2006qr}, 
the left-handed doublet $q_L$ and the singlet $t_R$ are embedded as
\begin{equation}
 \psi_L^t = {\Delta_L^t}^\dagger q_L \sim \mathbf{5}_{2/3}, \quad 
 \psi_R^t = {\Delta_R^t}^\dagger t_R \sim \mathbf{5}_{2/3},
\end{equation}
where the spurions
\begin{equation}
\begin{split}
\Delta_L^t & =  \frac{1}{\sqrt{2}} \begin{pmatrix} 0 & 0 & 1 & -i & 0 \\
 1 & i & 0 & 0 & 0 \end{pmatrix}
 \,,\quad 
 \Delta_R^t = - i \begin{pmatrix} 0 & 0 & 0 & 0 & 1
 \end{pmatrix}\,
\end{split}
\end{equation}
parameterize the $SO(5)$ breaking due to the fact that they do not fill complete $SO(5)$ multiplets.
In the proposed vector-like extension with soft Goldstone breaking, the sMCHM$_5$, we augment $\psi_L^t$ and $\psi_R^t$ to full $SO(5)$ representations by introducing two (vector-like) elementary 
SU(2)$_L$ doublets, $w$ and $v$, and a singlet, $s$, such that
\begin{equation}
 \psi_L^t = {\Delta_L^t}^\dagger q_L + \Delta_w^\dagger w_L 
 + \Delta_s^\dagger s_L, \quad 
 \psi_R^t = {\Delta_R^t}^\dagger t_R + \Delta_w^\dagger w_R
 +\Delta_v^\dagger v_R,
\end{equation}
where
\begin{equation}
\Delta_s = \Delta_R^t, \quad
\Delta_v = \Delta_L^t, \quad
\Delta_w =  \frac{1}{\sqrt{2}} \begin{pmatrix} 1 & -i & 0 & 0 & 0 \\
 0 & 0 & 1 & i & 0 \end{pmatrix}.
\end{equation}
The mass Lagrangian in the elementary sector reads
\begin{equation}
\label{eq:Lel}
\begin{split}
 -\mathcal{L}_\text{el} = & \, m_w (\bar{w}_L w_R + \bar{w}_R w_L)+
 m_v (\bar{v}_L v_R + \bar{v}_R v_L)+
 m_s ( \bar{s}_L s_R +  \bar{s}_R s_L)\\
 &
 + (m_1 \bar{s}_L t_R + m_2 \bar{q}_L v_R + \text{h.c.}) \,,
 \end{split}
\end{equation}
allowing to make the new fermions heavy via vector-like mass terms, and can 
be rewritten in terms of the conventional $\Delta_{L,R}^t$ spurions as
\begin{equation}\label{eq:Lel2}
\begin{split}
-\mathcal{L}_\text{el} = \ &
 m_w \bar{\psi}^t_L \psi^t_R 
  + m_v \bar{v}_L \Delta_L^t \psi_R^t
  + m_s \bar{s}_R \Delta_R^t \psi_L^t +\\
  &
 \left\{ (m_2 - m_w) \bar{q}_L \Delta^t_L \psi^t_R
 +(m_w + m_1) \bar{\psi}_L^t {\Delta_R^t}^* t_R +
   \,
 \text{h.c.} \,\right\} .
 \end{split}
\end{equation}
The elementary fields $\psi^t_L$ and $\psi^t_R$
formally mix with the composite resonances
as in the MCHM$_5$, i.e. (see e.g. \cite{Contino:2006qr,Panico:2011pw}),
	\begin{equation}	
	\begin{split}
	\label{eq:Lmm}
	 {\cal L}_{\rm mass} = 
	 & -m_Q \bar{Q}_L Q_R - \tilde m_T \bar{\tilde T}_L
	{\tilde T}_R   \\
	&  
	 -\,y^{}_L f\,  \bar{\psi}^t_{LI}
	\left( a_L U_{Ii} Q_R^i +  b_L U_{I5} \tilde T_R \right) \\
	&  -\,y^{}_R f\,  \bar{\psi}^t_{RI}
	\left( a_R U_{Ii} Q_L^i +  b_R U_{I5} \tilde T_L  \right)
	+{\rm h.c.}\,,
	\end{split}
	\end{equation}
where only the lightest top partners are kept
and $f$ is the Goldstone-Higgs decay constant.
As mentioned, this mixing does not break explicitly
$SO(5)$, while new sources of
$SO(5)$ breaking emerge in the elementary 
sector via the vector-like masses~\eqref{eq:Lel}.

We note that in the 5D holographic dual (see \cite{Agashe:2004rs,Contino:2003ve,Contino:2006qr}), our setup would correspond to choosing the same boundary conditions for {\it all} components of the fermionic bulk fiveplets of $SO(5)$ -- whose zero-modes 
contain the SM fermions -- thereby respecting $SO(5)$ at the first place.
The additional zero-modes that emerge due to these universal boundary conditions are then projected out via finite ($SO(5)$ breaking) vector-like boundary mass terms on the UV brane, instead of employing dedicated boundary conditions to remove them (as in the conventional approach), 
which would correspond to the limit of infinite vector-like masses.
This is similar to realizing EWSB via coupling to an IR-localized Higgs sector with a finite vev, instead of employing dirichlet boundary conditions to remove the massless modes of the weak gauge fields.

\begin{figure}[!t]
	\begin{center}

\begin{tikzpicture}
  \begin{feynman}
    \vertex (a) ;
    \vertex [above right=of a] (b);
    \vertex [below right=of b] (c);
    \vertex [below left=of c] (d);
    \diagram* {
      (a) -- [fermion, quarter left, edge label=\( s_R \), insertion=0.] (b) 
      -- [fermion, quarter left, edge label=\(\psi^t_L\), insertion=0.] (c) 
      -- [fermion, quarter left, edge label=\(\tilde T_R\), insertion=0.] (d) 
      -- [fermion, quarter left, edge label=\(\psi^t_L\), insertion=0.] (a) 
      };
    \draw [opacity = 0] (b) -- (b) node [opacity=1 , pos=0.5, above] {  \(\text{\footnotesize $m_s \Delta_R^t$}\)};
    \draw [opacity = 0] (c) -- (c) node [opacity=1 , pos=0.5, right] {  \(\text{\footnotesize $y_L U$}\)};
    \draw [opacity = 0] (d) -- (d) node [opacity=1 , pos=0.5, below] {  \(\text{\footnotesize $y_L U$}\)};  
    \draw [opacity = 0] (a) -- (a) node [opacity=1 , pos=0.5, left] {  \({\text{\footnotesize $m_s \Delta_R^t$}}\)};  
  \end{feynman} 
 \end{tikzpicture}
\quad \
\begin{tikzpicture}
  \begin{feynman}
    \vertex (a) ;
    \vertex [above right=of a] (b);
    \vertex [below right=of b] (c);
    \vertex [below left=of c] (d);
    \diagram* {
      (a) -- [fermion, quarter left, edge label=\( q_L \), insertion=0.] (b) 
      -- [fermion, quarter left, edge label=\(\psi^t_R\), insertion=0.] (c) 
      -- [fermion, quarter left, edge label=\(\tilde T_L\), insertion=0.] (d) 
      -- [fermion, quarter left, edge label=\(\psi^t_R\), insertion=0.] (a) 
      };
    \draw [opacity = 0] (b) -- (b) node [opacity=1 , pos=0.5, above] {  \(\text{\footnotesize $\delta_m \Delta^t_L$}\)};
    \draw [opacity = 0] (c) -- (c) node [opacity=1 , pos=0.5, right] {  \(\text{\footnotesize $y_R U$}\)};
    \draw [opacity = 0] (d) -- (d) node [opacity=1 , pos=0.5, below] {  \(\text{\footnotesize $y_R U$}\)};  
    \draw [opacity = 0] (a) -- (a) node [opacity=1 , pos=0.5, left] {  \({\text{\footnotesize $\delta_m \Delta^t_L$}}\)};  
  \end{feynman} 
 \end{tikzpicture}
 \quad\quad
 \begin{tikzpicture}

  \begin{feynman}
    \vertex (a) ;
    \vertex [above right=of a] (b);
    \vertex [below right=of b] (c);
    \vertex [below left=of c] (d);
    \diagram* {
      (a) -- [fermion, quarter left, edge label=\(q_L\)] (b) 
      -- [quarter left, edge label=\(y_L \Delta_L^t U\), insertion=0.5] (c) 
      -- [fermion, quarter left, edge label=\(\tilde T_R\)] (d)
      -- [quarter left, edge label=\( y_L \Delta_L^t U \), insertion=0.5] (a)  
      };
      \end{feynman}
   \end{tikzpicture}
	\caption{\label{fig:diags}
	Two possible contributions to the Higgs potential in the sMCHM$_5$ 
	(left), where $\delta_m \equiv m_2 - m_w$, compared to the MCHM$_5$ (right).}
	\end{center}
\end{figure}
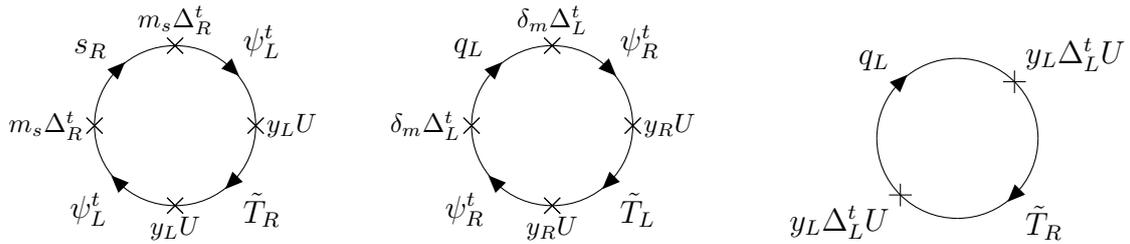

The different $SO(5)$ breaking results in a modified spurion 
structure for the Higgs potential.
In the MCHM$_5$, the spurions were given by 
the linear mixings $y_{L,R} \Delta_{L,R}^t$, which always come together with the Goldstone
matrix $U$.
In the sMCHM$_5$, however, the $SO(5)$ breaking
is given by the vector-like masses in the elementary
sector, and thus an extra interaction is needed to make
the Goldstone matrix enter the loop. This is schematically
shown in Fig.~\ref{fig:diags} 
for two example contributions, corresponding to the third 
and fourth term in Eq.~(\ref{eq:Lel2}).
In the next section, we will turn to a quantitative 
analysis of the Higgs potential -- focusing on the interplay of 
the Higgs mass and the top-partner masses -- where we will in 
particular see that the latter can be considerably lifted in
the sGB setup, while in the limit of heavy vector-like masses the 
sMCHM$_5$ recovers the properties of the MCHM$_5$. 

\section{Explicit Results in Two-Site Model}\label{sec:2site}
We perform our quantitative analysis in a two-site implementation
of the sGB scenario, where only the first layer of resonances is kept
in the spirit of discrete models, see \cite{Panico:2011pw,DeCurtis:2011yx}.
Such a setup has the major advantage of being simple and transparent, but
does not give a fully calculable Higgs potential, which is however
not required for our study. 
The two-site model is given explicitly by the Lagrangian \eqref{eq:Lmm},
with $a_L = a_R = b_L = b_R = 1$, together
with the elementary fields, whose dynamics is specified by
\eqref{eq:Lel2}.

The Coleman-Weinberg potential for the Higgs field is given by
\begin{equation}\label{eq:CW}
 V(h) = - \frac{2 N_c}{8 \pi^2}
 \int_0^\infty \text{d}p \, p^3 \, \text{ln}
 \left\{\text{det}[p^2 \mathbb{1} + m^\dagger m (h)] \right\},
\end{equation}
where $m(h)$ is the field-dependent fermion mass
matrix, such that
\begin{equation}
\text{det}[p^2 \mathbb{1} + m^\dagger m (h)]=
 1 + a(p^2)\, \text{sin}^2(h/f) + 
 b(p^2)\, \text{sin}^4(h/f),
\end{equation}
where $a(p^2)$ and $b(p^2)$ are functions of the fermion masses.
Expanding the logarithm up to $\text{sin}^4(h/f)$, one finds
\begin{equation}
 V(h) =  \alpha \, \text{sin}^2(h/f)
 + \beta \, \text{sin}^4(h/f),
\end{equation}
leading to the Higgs mass
\begin{equation}\label{eq:higgsmass}
 m_h^2 = 2 \beta/f^2 \, \text{sin}^2(2 v/f),
\end{equation}
where 
\begin{equation}\label{eq:alfabeta}
 \alpha = - \frac{2 N_c}{8 \pi^2} \int \text{d}p \, p^3 a(p^2), 
 \quad \beta = - \frac{2 N_c}{8 \pi^2} \int \text{d}p \, p^3 
 \left(b(p^2) - \frac{a^2(p)}{2}\right).
\end{equation}
We find that $\alpha$ diverges logarithmically and $\beta$ 
is finite at the order $\mathcal{O}(y_{L,R}^4)$, as in 
the two-site MCHM$_5$.
The divergence of $\alpha$ can be cured introducing
a renormalization scale $\mu$, which is fixed to
reproduce the correct Higgs vaccum expectation value (vev). 
As a consequence, $v \lesssim f$ does not require  $a(p^2) \lesssim b(p^2)$,
so that $a^2(p^2)$ and $b(p^2)$
can, \text{a priori},
equally contribute to $\beta$.
Notice that, despite the divergence in $\alpha$,
the relation between the Higgs mass and the fermion masses
is still predictable in the two-site model in terms of $\beta$.

In the MCHM$_5$, the top quark mixes with the 
doublet within $Q$ sharing the quantum numbers of $q_L$, denoted by $T$,
and with $\tilde{T}$,
the singlet with $t_R$ quantum numbers.
In the sMCHM$_5$, there are two corresponding doublets, 
identified with the two superpositions of 
$Q$ and the elementary doublet $v$, which we denote by $T_+$ and $T_-$. 
Similarly, $\tilde{T}$ splits into two states, made out of
$\tilde{T}$ and $s$ and referred to as $\tilde{T}_+$ and 
$\tilde{T}_-$. As a consequence,
the expression for the top mass is modified in the 
sMCHM$_5$, reading
\begin{equation}
\label{eq:mtop}
 m_t^2 = y_L^2 y_R^2 f^4 
 \frac{m_s^2 m_v^2 (m_Q - \tilde{m}_T)^2}{8 m_{T_+}^2 m_{T_-}^2 
 m_{\tilde{T}_+}^2 m_{\tilde{T}_-}^2}\text{sin}^2(2 v/f).
\end{equation}
In the limit of large $m_s$ and $m_v$, one can check that the elementary
fields effectively decouple and Eq.~\eqref{eq:mtop} 
approaches the result of the MCHM$_5$ \cite{Matsedonskyi:2012ym}, i.e.
\begin{equation}
 m_t^2 = y_L^2 y_R^2 f^4 
 \frac{(m_Q - \tilde{m}_T)^2}{8 m_T^2
 m_{\tilde{T}}^2 }
 \text{sin}^2(2 v/f).
\end{equation}

Since the vector-like masses for the different elementary fermion species,
$s, v$, and $w$, are in general independent quantities,
we divide our analysis in two parts.
In Sec.~\ref{sec:s}
we assume that only one vector-like fermion
is active below the cutoff scale 
$4 \pi f \simeq 10 \, \text{TeV}$
for $\xi = v^2/f^2 = 0.1$, 
while the other two are much heavier.
The case in which all the elementary states
are active, on the other hand, is presented in Sec.~\ref{sec:d}.

\subsection{Hierarchies in the elementary sector}\label{sec:s}

We start considering the case in which
the singlet $s$ is much lighter than $v$ and $w$. 
The top mass is then given by
\begin{equation}\label{eq:singtopmass}
 m_t^2 =  y_L^2 y_R^2 f^4 
 \frac{m_s^2(m_Q - \tilde{m}_T)^2}{8 m_T^2
 [m_{\tilde{T}}^2 m_s^2 + (y_L y_R f^2 + m_1 \tilde{m}_T)^2]}
 \text{sin}^2(2 v/f),
\end{equation}
where $m_T^2 = m_Q^2 + y_L^2 f^2$,
$m_{\tilde{T}}^2 = \tilde{m}_T^2 + y_R^2 f^2$.
From \eqref{eq:singtopmass}, we see that
the value for $m_t$ in the sMCHM$_5$ is always smaller
than in the MCHM$_5$, which can be recovered
for $m_s \rightarrow \infty$. By itself,
this effect would drive
the top partners to be even lighter
than in the MCHM$_5$. However, 
this is changed when the modification in the 
Higgs potential is taken into account.

In the following, we focus on $\beta$, which is calculable
in the two-site model and determines the Higgs mass \eqref{eq:higgsmass}.
For simplicity, we consider from now on the case in which the mixing
between the chiral and the vector-like elementary
fermions is negliglible, namely $m_{1,2}=0$ in \eqref{eq:Lel}.
By looking at \eqref{eq:alfabeta},
one can see that the $a^2(p^2)$ term
always increases the value of $\beta$.
Its effect was found to be (accidentally)
small for the two-site MCHM$_5$ in \cite{Matsedonskyi:2012ym}.
In the two-site sMCHM$_5$, it can instead be sizeable 
when all the vector-like fermions are light.
However, if only one of them happens to be active 
below the cut-off,
as discussed in this section, the $a^2(p^2)$ term 
can be safely neglected.
Within this approximation, we find
\begin{equation}
 \beta \simeq
 y_L^2 y_R^2 f^4 (m_Q - \tilde{m}_T)^2 
 \int_0^\infty \text{d}p p^3
 \frac{(p^2 + \tilde{m}_T^2) m_s^2-p^2(p^2+ m_T^2)}
 {2 p^2 (p^2+m^2_{\tilde{T}_+})(p^2+m^2_{\tilde{T}_-})
 (p^2 + m_T^2)(p^2 + \tilde{m}_T^2)}.
 \end{equation}
Notice that, compared to the two-site 
MCHM$_5$, there is one more
propagator associated with the splitting of
$\tilde{T}$ to $\tilde{T}_\pm$.
 
In the limit of $y_{L,R} f$ much smaller than the mass
of the vector-like fermions, we approximate the top mass as
\begin{equation}\label{eq:mtapp}
 m_t^2 \simeq \frac{y_L^2 y_R^2 f^4}{8 m_Q^2}
 (q-1)^2\, \text{sin}^2(2v/f),
\end{equation}
while $\beta$ reads
\begin{equation}\label{eq:betar}
 \beta(r^2) \simeq \frac{N_c}{16 \pi^2} y_L^2 y_R^2 f^4
 \frac{(1-q)^2}{1-q^2 r^2}
 \left[(r^2 + 1/q^2)F(q^2) - 2 F(r^2) \right],
\end{equation}
where $q=m_Q/\tilde{m}_T$, $r=\tilde{m}_T/m_s$,
and $F(x^2) = \frac{x^2}{1-x^2} \text{ln}\,\frac{1}{x^2}$.
One can show that
\begin{equation}\label{eq:gbr}
 \beta(r^2) - \beta(0) 
 = -\frac{N_c}{16 \pi^2} y_L^2 y_R^2 f^4
 (1-q)^2\frac{F(q^2) - F(1/r^2)}{q^2-1/r^2} \leq 0,
\end{equation}
where we have used $F(1/x^2) = F(x^2)/x^2$ and
$F'(x)>0$ for $x>0$.
The $\beta(0)$ term in \eqref{eq:gbr} corresponds
to the case in which the new singlet $s$ is infinitely heavy
and decouples. We checked that it coincides with
the conventional formula for $\beta$ in the two-site
MCHM$_5$, namely 
\begin{equation}\label{eq:beta0}
 \beta(0) \simeq \frac{ N_c}{16 \pi^2} y_L^2 y_R^2
 f^4 \frac{(1-q)^2}{q^2}F(q^2).
\end{equation}
As expected, Eq.\,\eqref{eq:gbr} shows that including the singlet $s$ 
always reduces the amount
of Goldstone breaking, leading to a lighter Higgs boson.
Combining \eqref{eq:mtapp} and
\eqref{eq:betar}, we find
\begin{equation}\label{eq:singhmass}
 m^2_Q = \frac{1}{16} y_L^2 y_R^2 f^2 \frac{(1-q)^2}{\beta(r^2)} 
 \left(\frac{m_h}{m_t}\right)^2,
\end{equation}
which relates the Higgs mass to the spectrum of resonances.
By inspecting \eqref{eq:betar}, we see 
that $\beta(r^2)$ vanishes for $q^2 r^4 = 1$
and it becomes negative for $q^2 r^4 > 1$,
the latter being in conflict with a viable electroweak symmetry breaking.
Approaching $q^2 r^4 = 1$, all the
fermions besides the top can be arbitrarily heavy while 
still keeping the Higgs light. 
However, such an extreme configuration corresponds
to infinite tuning in the parameter space
and thus it is never realized in a realistic scan, as we will see below, where the effects remain finite.
Nevertheless, this qualitative behavior
shows an important feature of the sGB.
In the MCHM$_5$, $\beta$ can be small
only for $q \rightarrow 1$, namely when $SO(5)_R$
is restored. 
This implies that the 
top mass would vanish in the same limit,
thus making this region disfavored.
In the sGB scenario, on the other hand, the breaking of the Goldstone symmetry is reduced
by the new degrees of freedom, see \eqref{eq:gbr}. This effect does not interfere much with the top mass,
Eq.\,\eqref{eq:mtapp} being indeed independent of $r$
to first approximation. 
Therefore, the Higgs potential and the top mass can be
disentangled such that a small $\beta$ does not necessarily imply 
a light top (which could be lifted only via ultra-light partners).

As an estimate for the mass of the lightest
eigenstate of the system, $m_l$, we take
\begin{equation}\label{eq:ml}
 m_l^2 \simeq \text{min}\{m_Q^2, \tilde{m}_T^2, m_s^2\}
 = m_Q^2 \times \text{min}\left\{1,\frac{1}{q^2},\frac{1}{q^2 r^2}\right\}
 = m_Q^2 \times \text{min}\left\{1,\frac{1}{q^2}\right\}
\end{equation}
where $m_Q^2$ is given in \eqref{eq:singhmass}. The estimate
above is correct up to mixing terms $\sim y_{L,R} f$.
The last equality in \eqref{eq:ml} is derived as follows.
By definition, $m_s$ can be the lightest state only for $r^2 \geq 1$
and $r^2 \geq 1/q^2$. Moreover, $r$ and $q$ must satisfy
$r^2 \leq r_0^2 = 1/|q|$, as discussed below \eqref{eq:singhmass}.
However, if $|q| > 1$ one immediately derives the contradiction $r^2 < 1$, while
if $|q| < 1$ one has $r^2 \geq 1/q^2 = r_0^4 > r_0^2$, which is 
clearly not compatible with $r^2 \leq r_0^2$. 
In conclusion, the spectrum favors a configuration with 
the elementary singlet $s$ not residing at the bottom. 
When the latter is about to become the lightest
state, $\beta$ flips sign and the electroweak vaccum is no longer a minimum.

\begin{figure}[t]
\centering
\includegraphics[width=7.3cm]{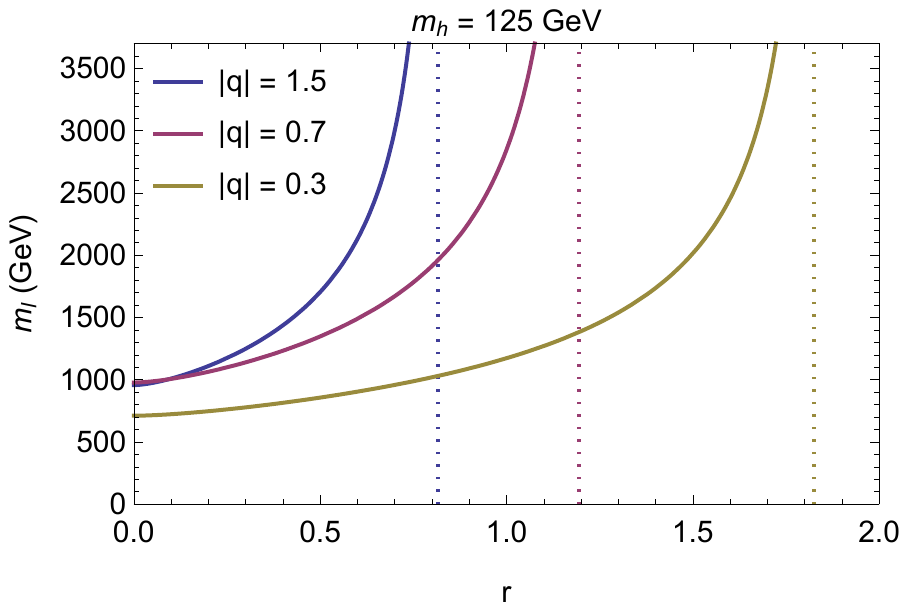}
\includegraphics[width=7.3cm]{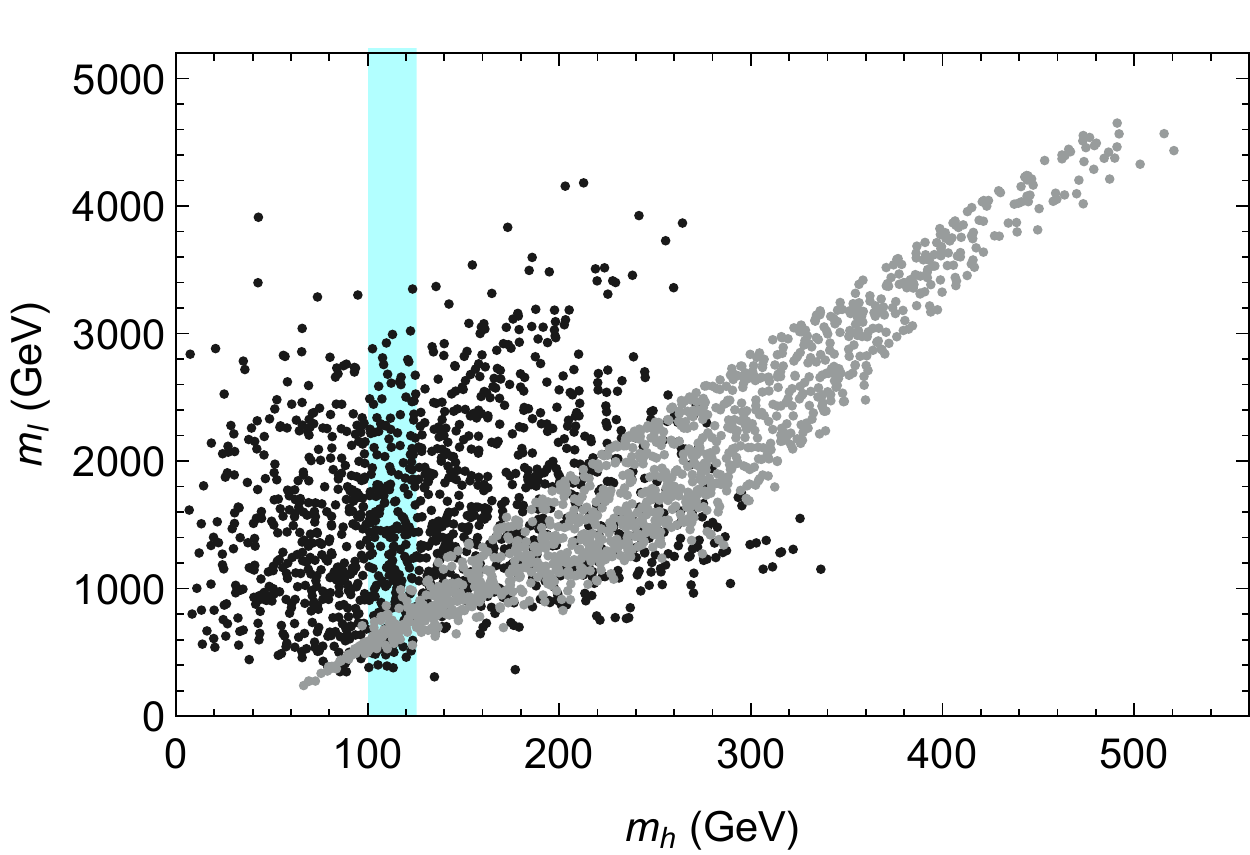}
\caption{\textbf{Left}: The mass of the lightest top partner
for different 
values of $|q|$ as a function of $r = \tilde{m}_T/m_s$
for $m_h = 125 \,\text{GeV}$. The dashed vertical lines indicate
the value of $r$ for which $q^2 r^4 = 1$. 
The result is the same for $r \rightarrow -r$.
\textbf{Right}: Scatter plot of $(m_h,m_l)$
for $0.5 \leq |r| \leq 2$ (black points) and $r=0$ (gray points).
We scan in the range $ 0.5 \leq |y_{L,R}| \leq 2$
and $0.5 \leq |\tilde{m}_T/\text{TeV}| \leq 4.5 $ 
such that $-2 \leq q \leq -0.3$.}
\label{fig:2}
\end{figure}

The value of $m_l$, as predicted by
\eqref{eq:ml}, is shown in 
the left panel of Fig.\,\ref{fig:2}
as a function of $r$ for 
different values of $q$.
We take $m_t \simeq 150 \, \text{GeV}$ for the top mass at the scale $f$.
The result of the MCHM$_5$ is recovered for $r=0$.
The mass of the lightest state
always increases with $r$
until it hits the bound $q^2 r^4 = 1$,
shown by the dotted vertical lines.
It is important to notice that $m_l$ can be
significantly heavier than in the standard case
for values of $r$ and $q$ which are far
from $q^2 r^4 = 1$. Thus, a heavier $m_l$
does not correspond to fine tuning in the parameter 
space but rather to a general prediction of the sMCHM$_5$.
This is confirmed by the results shown
in the right panel of Fig.\,\ref{fig:2},
where $(m_h,m_l)$ are obtained as an output 
from a numerical scan, using the exact expression for $\beta$
(including the $a^2(p^2)$ term in \eqref{eq:alfabeta}).
The window $100 \, \text{GeV} \leq m_h \leq 125 \, \text{GeV}$,
visualized by the blue stripe, is
considered to take into account running effects on
the actual value of the Higgs mass at the scale $f$.
The mass ratio $r$ is scanned
between $0.5\leq |r| \leq 2$ (black points).
Gray points correspond to the MCHM$_5$, i.e. $r=0$. 
The effect of the singlet $s$ can be seen
as effectively reducing
the Higgs mass which is consistent with a certain value of $m_l$.
For instance, $m_l \simeq 3 \, \text{TeV}$ is compatible
with $m_h \approx 100\, \text{GeV}$ in the sMCHM$_5$, whereas that would
require $m_h \gtrsim 300\, \text{GeV}$ in the conventional case.
This is just an equivalent way of looking at \eqref{eq:gbr}, 
where the ratio $\beta(r^2)/\beta(0) = m_h^2(r^2)/m_h^2(0) < 1$ gives the correct 
estimate for the compression of the Higgs spectrum.

Turning to the other cases, the situation in which only one of the doublets is light,
either $w$ or $v$, corresponds to a much more 
modest change with respect to the MCHM$_5$. 
In both cases, the $a^2(p^2)$ term in \eqref{eq:alfabeta}
is found to be small. The expression for $\beta$
driven by $w$ in isolation matches the conventional result
at the order $\mathcal{O}(y_{L,R}^4)$.
On the other hand, $v$ does reduce $\beta$ at the 
$\mathcal{O}(y_{L,R}^4)$, its effect being however more modest
due to an accidental factor $1/2$ in front of $F(r^2)$
(to be compared with a factor of $2$ in \eqref{eq:betar}):
\begin{equation}
 \beta(r^2) \simeq \frac{N_c}{16 \pi^2} y_L^2 y_R^2 f^4
 \frac{(1-q)^2}{1-q^2 r^2}
 \left[(1/q^2 -r^2/2)F(q^2) - F(r^2)/2 \right],
\end{equation}
where $q=m_Q/\tilde{m}_T$, as before,
while now $r=\tilde{m}_T/m_v$.
The fact that the contribution of the doublets is
modest in isolation does not mean that 
they are always negliglible.
In particular, the Goldstone symmetry can be in principle 
restored only if all the elementary vector-like states are
active. Of course, too light states are in conflict with the LHC searches,
the intermediate region being the topic of the next section.

\subsection{No hierarchies in the elementary sector}\label{sec:d}

We finally discuss the results for the
case in which no hierarchies are introduced
among the elementary vector-like fermions
$s$, $v$ and $w$. 
To this end, we derived an expression for $\beta$ following the same
procedure that lead to \eqref{eq:betar}, setting $m_v = m_w \equiv m_d$ for simplicity, which we refer to as $\beta_f$ in the following.
The relative spread between the elementary states
is parameterized by $x = m_d/m_s$, 
while $r = \tilde{m}_T/m_s$ relates
the elementary state $s$ to the composites.
We notice that the $a^2(p^2)$ term
in \eqref{eq:alfabeta} can actually 
be important for $x = \mathcal{O}(1)$.
The lightest state $m_l$ is now estimated as
\begin{equation}\label{eq:mld}
 m_l^2 \simeq \text{min}\{m_Q^2, \tilde{m}_T^2, m_s^2, m_d^2\}
 = m_Q^2 \times \text{min}\left\{1,\frac{1}{q^2},\frac{1}{q^2 r^2},
 \frac{x^2}{q^2 r^2} \right\}\,,
\end{equation}
while the top mass is still approximated by
\eqref{eq:mtapp}. 
The quantity $m_Q$ obeys \eqref{eq:singhmass},
where $\beta$ is now replaced by $\beta_f$.

We have checked that $\beta_f$ approaches zero for
$r \rightarrow \infty$, corresponding to massless elementary
fermions and thus restoring the Goldstone symmetry. 
In practice, such limit is not of much use
because it would introduce very light states in the spectrum.
Nevertheless, while approaching this limit, $\beta_f$ can have a zero also
at an intermediate value of $r$ depending on $x$ and $q$.
Although this is never realized
in a realistic scan, the possibility
of getting close to it without affecting
the top mass again shows the new feature of the 
sGB. 

We identify three different regions depending on the structure in the elementary sector, i.e., on the value of $x$.
For $|x| \leq 2$, the improvement in the sMCHM$_5$ is modest
and $m_l$ can be roughly $2 \, \text{TeV}$ at most.
The case of a large $|x|$ actually corresponds
to the singlet case discussed at the beginning
of Sec.\,\ref{sec:s}. Thus, we here focus 
on the intermediate case $2 \leq |x| \leq 4$.

\begin{figure}[t]
\centering
\includegraphics[width=7.31cm]{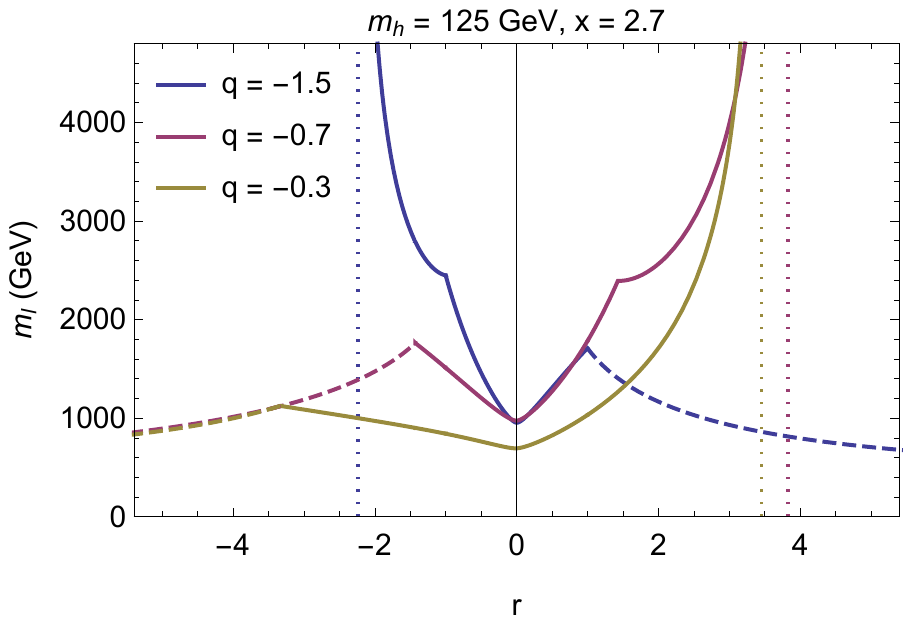}
\includegraphics[width=7.3cm]{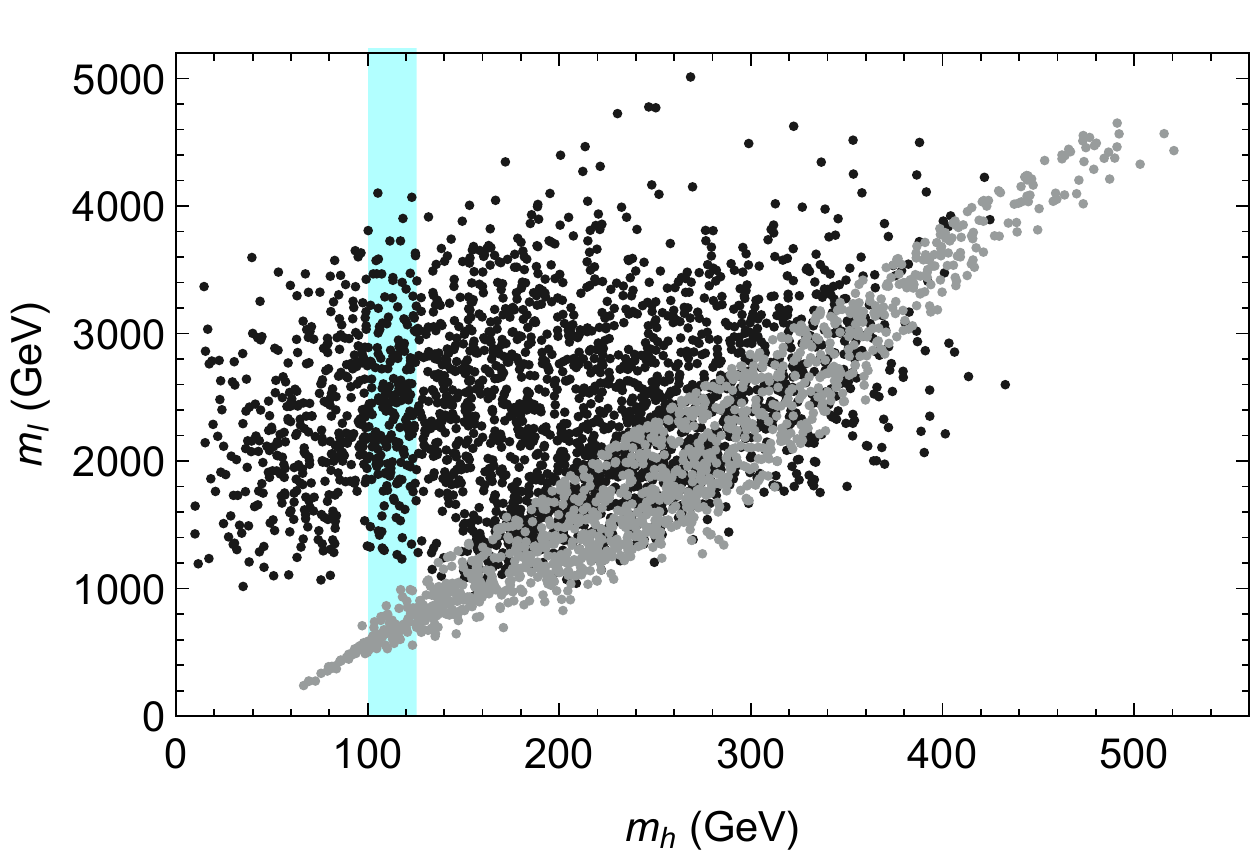}
\caption{\textbf{Left}: Analytical approximation for the mass of the lightest top partner in 
the case $m_w = m_v = m_d$ 
for  $q = (-1.5,-0.7,-0.3)$ as a function of $r = \tilde{m}_T/m_s$
with $m_h = 125 \,\text{GeV}$ and $x = m_d/m_s = 2.7$.
The vertical dotted lines indicate $r$
such that $\beta_f = 0$ for the different values of $q$.
The dashed curves show configurations for which 
the decrease in $m_l$ is driven by the lightness of the new singlet.
\textbf{Right}: The mass of the lightest top partner, $m_l$,
as a function of $m_h$ in the sMCHM$_5$
(black points) and in the MCHM$_5$ (gray points). 
The scan for the sMCHM$_5$ assumes
$1 \leq |y_{L,R}|\leq 2$
and $-2 \leq q \leq -0.3$. Moreover, we consider
$5 \, \text{TeV} \leq |\tilde{m}_T| 
\leq 10 \, \text{TeV}$, while
$ 1.5 \leq |r| \leq 5$. The doublet
masses $m_{w,v}$ are scanned independently in the range $2 \leq |m_{w,v}/m_s| \leq 4$. 
The scan for the MCHM$_5$ is the same as in Fig.\,\ref{fig:2}.
All the mass eigenstates
reside below $15 \, \text{TeV}$.}
\label{fig:comp}
\end{figure}

The analytical prediction based on $\beta_f$ is shown 
in the left panel of Fig.\,\ref{fig:comp} for $x=2.7$.
The knees signal that the elementary singlet
becomes the lightest state.
This happens for $r$ such that 
$r^2 \geq r^2_q = \text{max}(1,1/q^2)$.
Notice that this was not possible
in the case in which the $w$ and $v$ doublets
are infinitely heavy, 
since it would give a negative value of $\beta$,
as discussion below \eqref{eq:ml}.
The dotted lines show the location of $r$ such that
$\beta_f$ formally vanishes.
The values of $r$ above the
dotted lines actually lead to a viable
$m_l$, its value being just too large
to be shown in the plot. 
Although such high masses are
cut-off in a realistic scenario, 
this shows that the range of $r$ leading to a viable 
electroweak symmetry breaking is 
enlarged with respect to the singlet case
(see, e.g, the left panel of Fig.\,\ref{fig:2}).

Since the largest $m_l$ is typically found above the knee,
a heavy top partner favors the case of the singlet
$s$ as the lightest particle.
This implies that the spectrum can be stabilized
without requiring the spin-$1/2$ resonances
to lie much below the naive cutoff of
the strong dynamics, as needed in the conventional case to
reproduce the correct Higgs mass.

We explore this region of the parameter
space in the right panel of Fig.\,\ref{fig:comp},
where $(m_h,m_l)$ are obtained after a numerical scan.
We assume $\tilde{m}_T$ to be in the range 
$5 - 10\, \text{TeV}$, where the latter
coincides with the cutoff scale $4 \pi f$,
for $\xi = 0.1$.
Moreover, we scan $1.5 \leq |r| \leq 5$, 
while the doublet masses are independently scanned in the range $2 \leq |m_{v,w}/m_s| \leq 4$. 
To obtain the correct top mass,
we consider $1 \leq |y_{L,R}|\leq 2$. 
As we can see, the lightest
fermion state can now be pushed above $3.5 \, \text{TeV}$
in the blue band corresponding to $100 \, \text{GeV} \lesssim m_h 
\lesssim 125 \, \text{GeV}$.

\section{Benchmark points}\label{sec:BM}
We now choose two benchmark points to show how the spectrum
looks like, in the case in which only the elementary singlet
vector-like fermion $s$ is below the cutoff, 
presented at the beginning of Sec.\,\ref{sec:s},
and in the case of Sec.\,\ref{sec:d} where all the elementary
states are active.

\begin{figure}[t]
\centering
\includegraphics[width=2.4cm]{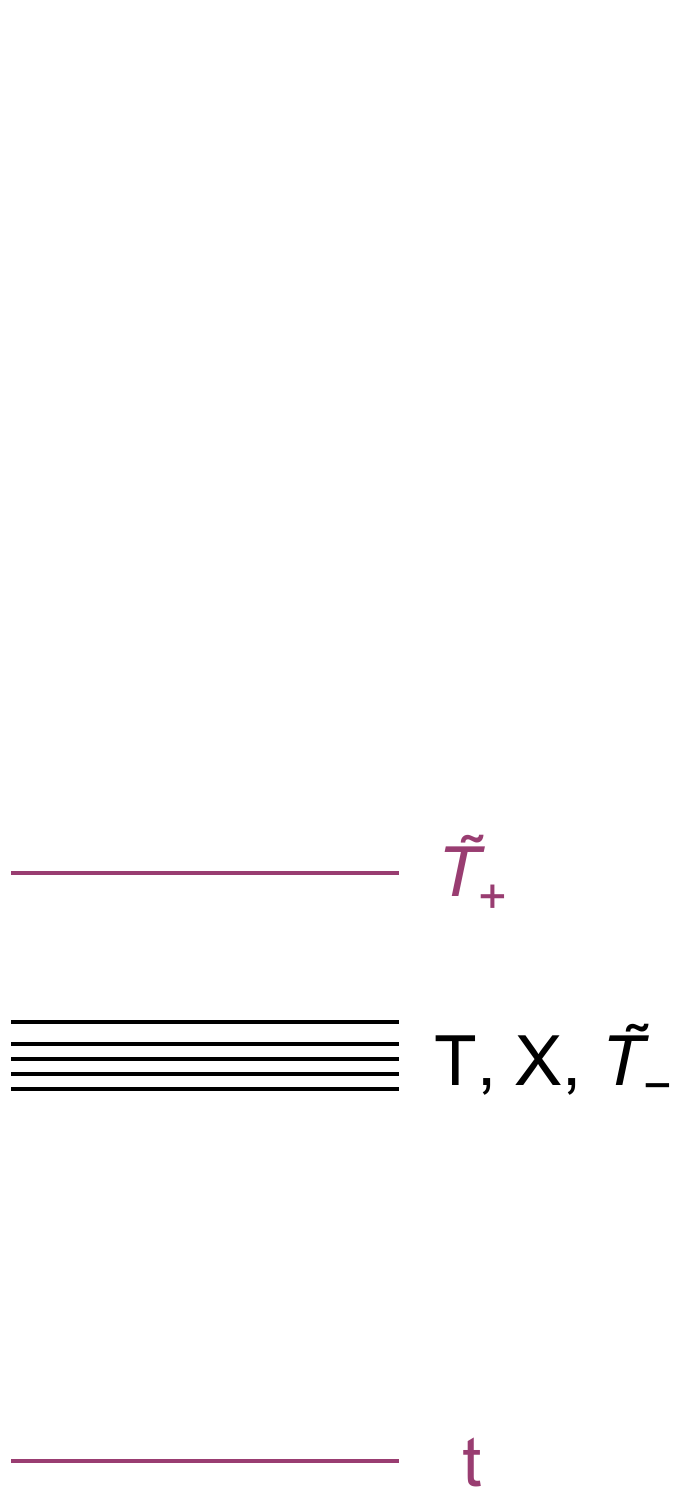}
\hspace{0.6cm}
\includegraphics[width=1.2cm]{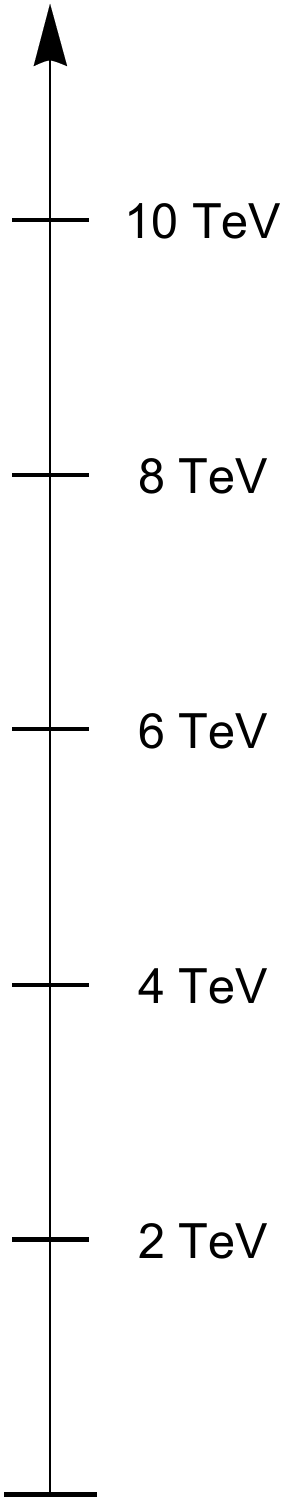}
\hspace{0.6cm}
\includegraphics[width=2.2cm]{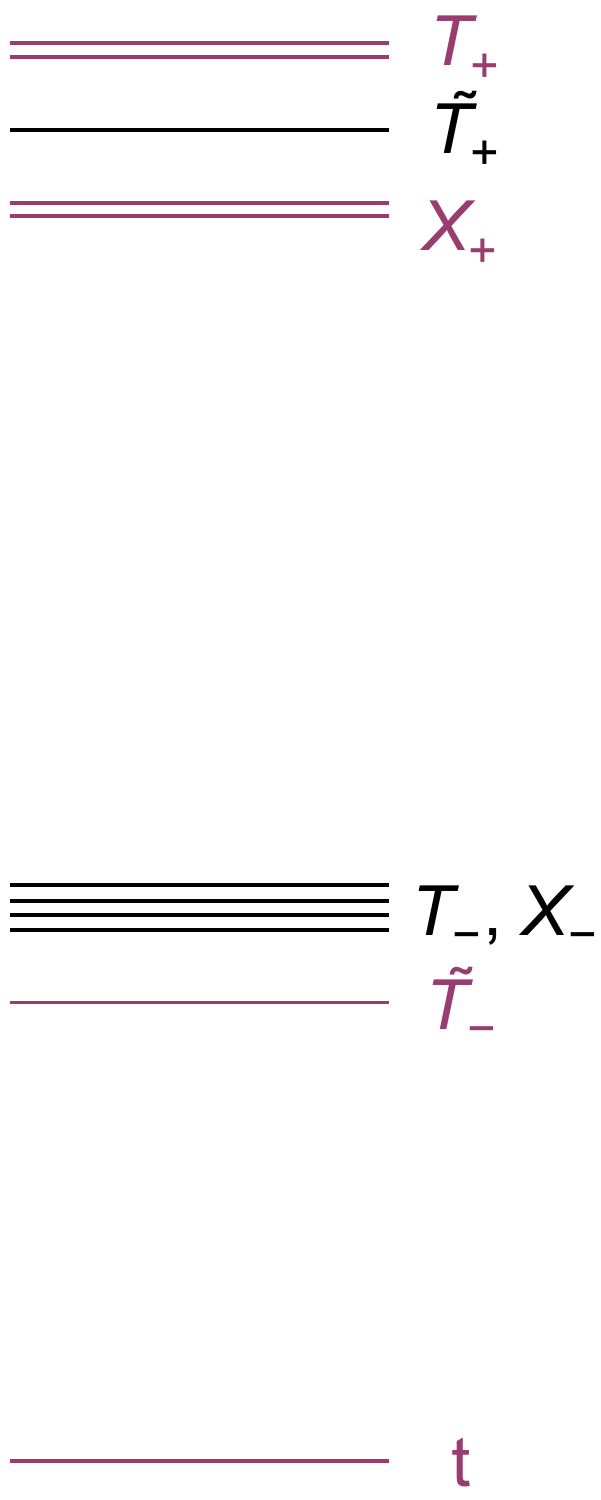}
\caption{\textbf{Left}: The spectrum as resulting
from the benchmark point $B_s$.
The lightest particle mainly overlaps
with the composite singlet $\tilde{T}$, with a mass
$m_l \simeq 2.7 \, \text{TeV}$. The heaviest state
mainly comes from the elementary singlet $s$ with a mass 
$\sim 4 \, \text{TeV}$.
The states colored by black are mainly composites, whereas
the purple ones are mainly elementary.
We have denoted by $X$ the doublet other than $T$, forming the
$SO(4)$ fourplet $Q$.
\textbf{Right}: The spectrum as resulting
from the benchmark point $B_f$.
The lightest particle is mainly
the elementary singlet $s$, with a mass
$m_l \simeq 3.4 \, \text{TeV}$.
We have denoted by $X_{\pm}$ the eigenstates resulting
from the mixing of $X$ and $w$.
Compared to the left panel, the effect of bringing
down the doublets $T_+$ and $X_+$ is to 
lift the lightest state together with 
the composite resonaces, which are found at $9.3 \, \text{TeV}$
(singlet of $SO(4)$) and at $4 \, \text{TeV}$ (fourplet of $SO(4)$).}
\label{fig:ben}
\end{figure}

For the singlet case we take $B_s$ as
\begin{equation}
B_s: \left\{y_L = 1.4, y_R = 1.3, \tilde{m}_T = 3 \, \text{TeV},
m_s = 3.8 \, \text{TeV}\right\},
\end{equation}
so that $r =\tilde{m}_T/m_s \simeq 0.8$,
$q = -0.9$ and $m_t \simeq 140 \, \text{GeV}$.
Notice that $q^2 r^4 \simeq 0.3$, which is far from
the (approximate) zero of $\beta$ at $q^2 r^4 = 1$.
The lightest eigenstate is the mainly composite singlet state
$\tilde{T}_-$, with a mass $m_l \simeq 2.7 \, \text{TeV}$.
The Higgs mass is found to be $m_h \simeq 110 \, \text{GeV}$.
The spectrum is shown in the left panel of Fig.\,\ref{fig:ben}. 
The states coloured black correspond to mainly composite states,
whereas the purple ones are mostly elementary.
The composite resonaces lie in the range $2.7-3.0 \, \text{TeV}$.
The heaviest state is the mainly elementary singlet $\tilde{T}_+$
with a mass around $4 \, \text{TeV}$.

For the case discussed in Sec.\,\ref{sec:d}, where all the elementary
vector-like fermions are kept in the spectrum, 
we take $B_f$ to be
\begin{equation}\begin{split}
 B_f: \{ 
 y_L = 1.8, & y_R = 1.8,  x = 2.8,
y = 2.5, \tilde{m}_T = 9 \, \text{TeV}, 
m_s = 3.5 \, \text{TeV} \},
\end{split}
\end{equation}
which yields 
$q \simeq - 0.4$ and $m_t \simeq 140 \, \text{GeV}$.
The lightest state is now the mostly elementary
singlet $\tilde{T}_-$ with $m_l \simeq 3.4 \, \text{TeV}$.
The Higgs mass is $m_h \simeq 120 \, \text{GeV}$.
The spectrum is given in the right panel of Fig.\,\ref{fig:ben},
where the same color convention is used to distinguish 
the mostly elementary states from the mainly composites.
The overall effect of bringing the doublets
$v$ and $w$ down with respect to left panel (where they are decoupled)
is to lift the lightest state, which is now mostly elementary.
This opens the possibility of having the spin-$1/2$
resonaces of the strong dynamics
closer to the cutoff $\Lambda \simeq 10 \, \text{TeV}$,
in this example $m_{\tilde{T}_+} \simeq 9.3 \, \text{TeV}$
and $m_{T_-,X_-} \simeq 4 \, \text{TeV}$.

\section{The general case of vector-like
elementary fermions}
\label{sec:Gen}
\begin{figure}[t]
\centering
\includegraphics[width=7.7cm]{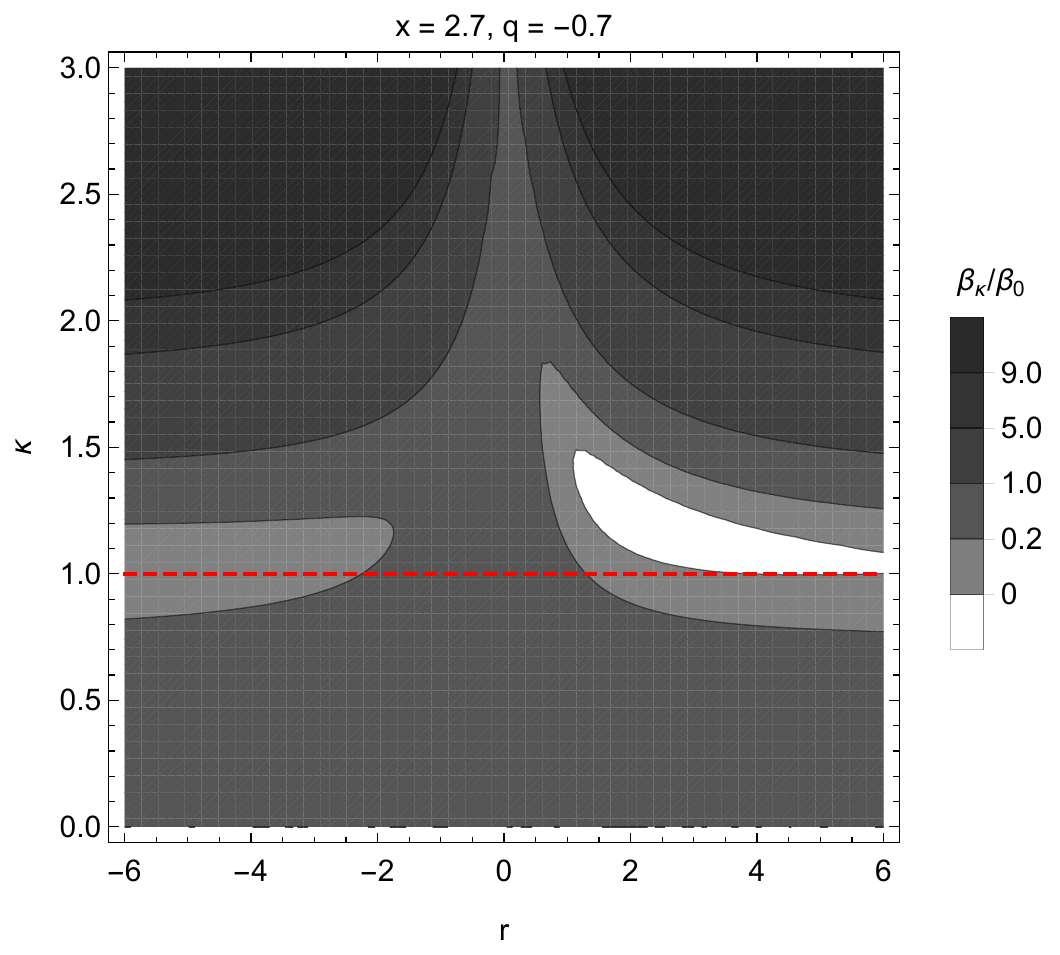}
\caption{The ratio $\beta_\kappa/\beta_0$ as a contour plot
in the $(r,\kappa)$ plane for $x=2.7$ and $q = -0.7$. 
The plot is mirrored for $\kappa \rightarrow -\kappa$. See text for details.} 
\label{fig:pk}
\end{figure}
The setup discussed in Sec.\,\ref{sec:setup}
is based on the fact that the new vector-like
fermions are embedded together with the chiral
fermions to form complete $SO(5)$ representations.
Therefore, partial compositeness now preserves
the global symmetry and its breaking is 
sequestered to non-zero masses
for the vector-like fermions.

We now relax this assumption and show what happens if the new fermions can couple arbitrarily to the composite resonances.
To do so, we recompute $\beta$ perturbing the 
partial compositeness couplings of $s$, $v$ and $w$ by a common
factor $\kappa$ (now referred to as $\beta_\kappa$), such that $\kappa=0$
reproduces the MCHM$_5$ and $\kappa=1$ corresponds to the analysis of Sec.\,\ref{sec:d} for the sMCHM$_5$.

The ratio $\beta_\kappa/\beta_0$ measures the degree
of Goldstone breaking and is shown via a contour plot 
in Fig.\,\ref{fig:pk}, where $r = \tilde{m}_T/m_s$, $x = m_d/m_s = 2.7$,
with $m_d$ the degenerate mass of the doublets $v$ and $w$,
for a reference value of $q = m_Q/\tilde{m}_T = -0.7$.
As we can see, increasing $\kappa$ from zero helps 
reducing the Goldstone symmetry breaking,
up to the point $\kappa \gtrsim 2$, from which on it is enhanced.
The optimal region is indeed around $\kappa^2 \approx 1$,
namely the $SO(5)$ symmetric point for partial compositeness.
Thus, organizing the new elementary fermions to achieve
complete $SO(5)$ representations (when possible)
is not only motivated by symmetry arguments (see also the discussion on 
the holographic picture in Sec.\,\ref{sec:setup})
but it turns out to be in general the safest option to avoid light partners.

\section{Conclusions}
\label{sec:conc}

We have proposed a new way of breaking the Goldstone symmetry in 
CH scenarios, which is responsible for a non-vanishing Higgs potential.
Instead of violating it via assuming the (SM-like) fermions not to fill
complete representations of the global symmetry
($SO(5)$ in our case), we break it 'softly', i.e. via finite vector-like
masses lifting the additional degrees of freedom that fill the
representations beyond direct LHC reach (opposed to fully eliminating 
them from the spectrum). As we have shown explicitly for a two-side incarnation, 
this allows to reduce the amount of Goldstone-symmetry breaking 
such that the large top mass can be reproduced (for fixed $f$)
without the necessity of vastly lowering the masses of the lightest 
top partners, the latter option starting to be in tension with LHC searches. 
For example, for $f \simeq 780 \, \text{GeV}$ it is possible to lift the lightest
partners from $m_l\lesssim 1\,$TeV up to $m_l\sim 4\,$TeV, coming closer 
to the general scale of CH resonances. While the light top partners might 
thus be hard to detect directly at the LHC (and are generically above current LHC limits),
for this setup they would be fully discovereable at the FCC.
In this context, we finally note that a further phenomenological survey of the
proposed sMCHM would be interesting, including an analysis of electroweak 
precision observables, Higgs physics, and direct searches for (elementary and composite) 
fermionic resonances, which is however beyond the scope of the present article.

\section*{Acknowledgments}

We are grateful to Roberto Contino and Tommi Alanne for useful discussions.

\bibliographystyle{unsrt}
\bibliography{refs}

\begin{thebibliography}{10}

\bibitem{Kaplan:1983fs}
David~B. Kaplan and Howard Georgi.
\newblock {SU(2) x U(1) Breaking by Vacuum Misalignment}.
\newblock {\em Phys. Lett.}, 136B:183--186, 1984.

\bibitem{Kaplan:1983sm}
David~B. Kaplan, Howard Georgi, and Savas Dimopoulos.
\newblock {Composite Higgs Scalars}.
\newblock {\em Phys. Lett.}, 136B:187--190, 1984.

\bibitem{Dugan:1984hq}
Michael~J. Dugan, Howard Georgi, and David~B. Kaplan.
\newblock {Anatomy of a Composite Higgs Model}.
\newblock {\em Nucl. Phys.}, B254:299--326, 1985.

\bibitem{Kaplan:1991dc}
David~B. Kaplan.
\newblock {Flavor at SSC energies: A New mechanism for dynamically generated
  fermion masses}.
\newblock {\em Nucl. Phys.}, B365:259--278, 1991.

\bibitem{Agashe:2004rs}
Kaustubh Agashe, Roberto Contino, and Alex Pomarol.
\newblock {The Minimal composite Higgs model}.
\newblock {\em Nucl. Phys.}, B719:165--187, 2005.

\bibitem{Contino:2003ve}
Roberto Contino, Yasunori Nomura, and Alex Pomarol.
\newblock {Higgs as a holographic pseudoGoldstone boson}.
\newblock {\em Nucl. Phys.}, B671:148--174, 2003.

\bibitem{Contino:2006qr}
Roberto Contino, Leandro Da~Rold, and Alex Pomarol.
\newblock {Light custodians in natural composite Higgs models}.
\newblock {\em Phys. Rev.}, D75:055014, 2007.

\bibitem{Matsedonskyi:2012ym}
Oleksii Matsedonskyi, Giuliano Panico, and Andrea Wulzer.
\newblock {Light Top Partners for a Light Composite Higgs}.
\newblock {\em JHEP}, 01:164, 2013.

\bibitem{Pomarol:2012qf}
Alex Pomarol and Francesco Riva.
\newblock {The Composite Higgs and Light Resonance Connection}.
\newblock {\em JHEP}, 08:135, 2012.

\bibitem{Csaki:2008zd}
Csaba Csaki, Adam Falkowski, and Andreas Weiler.
\newblock {The Flavor of the Composite Pseudo-Goldstone Higgs}.
\newblock {\em JHEP}, 09:008, 2008.

\bibitem{DeCurtis:2011yx}
Stefania De~Curtis, Michele Redi, and Andrea Tesi.
\newblock {The 4D Composite Higgs}.
\newblock {\em JHEP}, 04:042, 2012.

\bibitem{Goertz:2018dyw}
Florian Goertz.
\newblock {Composite Higgs theory}.
\newblock {\em PoS}, ALPS2018:012, 2018.

\bibitem{Carmona:2014iwa}
Adrian Carmona and Florian Goertz.
\newblock {A naturally light Higgs without light Top Partners}.
\newblock {\em JHEP}, 05:002, 2015.

\bibitem{Carmona:2015ena}
Adrian Carmona and Florian Goertz.
\newblock {Lepton Flavor and Nonuniversality from Minimal Composite Higgs
  Setups}.
\newblock {\em Phys. Rev. Lett.}, 116(25):251801, 2016.

\bibitem{Carmona:2016mjr}
Adrian Carmona and Florian Goertz.
\newblock {A flavor-safe composite explanation of $R_K$}.
\newblock {\em Nucl. Part. Phys. Proc.}, 285-286:93--98, 2017.

\bibitem{Carmona:2017fsn}
Adrián Carmona and Florian Goertz.
\newblock {Recent $\boldsymbol{B}$ Physics Anomalies - a First Hint for
  Compositeness?}
\newblock 2017.

\bibitem{Panico:2012uw}
Giuliano Panico, Michele Redi, Andrea Tesi, and Andrea Wulzer.
\newblock {On the Tuning and the Mass of the Composite Higgs}.
\newblock {\em JHEP}, 03:051, 2013.

\bibitem{Panico:2011pw}
Giuliano Panico and Andrea Wulzer.
\newblock {The Discrete Composite Higgs Model}.
\newblock {\em JHEP}, 09:135, 2011.

\end{thebibliography}

\end{document}